\documentclass[12pt]{article}
\pdfoutput=1
\usepackage[a4paper]{geometry}
\usepackage{amsmath,amssymb,amsfonts,latexsym,braket,amsthm,datetime}
\usepackage{graphicx}  
\usepackage[colorlinks=true, pdfstartview=FitV, linkcolor=blue, citecolor=blue, urlcolor=blue]{hyperref}
\newcommand {\tb}{ }

\newcommand{\ep}{\qquad {\vrule height 10pt width 8pt depth 0pt}}
\newcommand {\mD}{{\bf D}}

\newcommand {\bR}{{\mathbb R}}
\newcommand {\bN}{{\mathbb N}}

\newcommand {\bZ}{{\mathbb Z}}
\newcommand {\bI}{{\mathbb I}}

\newcommand {\bA}{{\mathbb A}}

\newcommand {\bC}{{\mathbb C}}

\newcommand {\bH}{{\mathbb H}}
\newcommand {\bT}{{\mathbb T}}
\newcommand {\bP}{{\mathbb P}}

\newcommand {\bV}{{\mathbb V}}

\newcommand {\cH}{{\mathcal H}}

\newcommand {\cO}{{\mathcal O}}
\newcommand {\cS}{{\mathcal S}}
\newcommand {\cT}{{\mathcal T}}

\newtheorem{theorem}{Theorem} [section]
\newtheorem{lemma}[theorem]{Lemma}
\newtheorem{propo}[theorem]{Proposition}
\newtheorem{definition}[theorem]{Definition}

\newtheorem {remark}[theorem]{Remark}
\newtheorem {remarks}[theorem]{Remarks}

\newtheorem*{hypo}{Hypothesis}
\begin{document}
\title{Energy-time uncertainty principle and lower bounds on sojourn time}
\author{
Joachim Asch,\thanks{Universit\'e de Toulon, CNRS, CPT UMR 7332,  CS 60584, F--83041 Toulon cedex 9, France}\ 
                        \thanks{Aix Marseille Universit\'e, CNRS, CPT UMR 7332, F--13288 Marseille cedex 9, France}\ 
Olivier Bourget, Victor Cort\'es, Claudio Fernandez\thanks{
Departamento de Matem\'aticas,Pontificia Universidad Cat\'olica de Chile, Av. Vicu\~{n}a Mackenna 4860, C.P. 690 44 11, Macul, Santiago, Chile}
}
\date{18.01.16}
\maketitle

\abstract{One manifestation of quantum resonances is a large  sojourn time, or autocorrelation,  for states which are initially localized. We elaborate on Lavine's time-energy uncertainty principle and give an estimate on the sojourn time.   For the case of perturbed embedded eigenstates the bound is explicit and involves  Fermi's Golden Rule. It is valid for a very general class of systems.  We illustrate the theory by applications  to resonances for  time dependent systems including the AC Stark effect as well as  multistate systems.}

\section{Introduction}

By a state in a Hilbert space $\cH$ we understand a normalized vector $\psi\in\cH$, respectively the associated projector $P=\ket{\psi}\bra{\psi}$. Given a selfadjoint operator $H$ in $\cH$ and the dynamics  generated by $H$, the sojourn time for a state $\psi $ is defined by

\begin{equation}
\mathcal{T}=\mathcal{T} (H, \psi) :=  \int_{-\infty}^{\infty} | \langle
 \psi, {\large{e}}^{-iHt}\psi \rangle |^2 \, dt \, .
\end{equation}
$\mathcal{T}$ is a  measure of the total amount of time the system spends in its initial state; it equals the autocorrelation $\int_{-\infty}^{\infty}  tr\left(P P_{t}\right) \, dt \,$, where $P_{t}:={\large{e}}^{-iHt}P{\large{e}}^{iHt}$; it is infinite if $\psi$ is an eigenstate of $H$. The main result we shall prove in this paper is a lower bound on $\cT$ for a perturbed embedded eigenstate $\psi$. 

More precisely, consider 
\[H=H_{0}+\kappa V_\kappa, \]
  $E_{0},\psi$ such that $H_{0}\psi=E_{0}\psi$ and $E_{0}$ embedded in the continuous spectrum of $H_{0}$, $\kappa$ a small parameter.  \tb{We assume that the eigenvalue is simple and that the restriction of $H_{0}$  to $Ran P^{\perp}$  has good propagation properties which persist upon perturbation by $\kappa V_\kappa$; it is known that this is implied by a Mourre estimate which we assume in hypothesis (A) below. It involves, in particular, multiple commutators of unbounded operators;  we now state and discuss our main result and refer the reader to  Section \ref{sec:qlb} for a precise definition of the mathematically more involved objects.   Denote $\chi(H\in I)$ the spectral projection of a selfadjoint operator $H$ on a Borel set $I$,
$ad_{A}(B)=\lbrack A,B\rbrack=AB-BA$  the commutator, and  by $ad_{A}^k(B)=ad_{A}^{k-1}(ad_A(B))$ its iterations. }

We assume that the quadruple $H_{0}, V, A,I$ satisfies

\begin{hypo}[${\bf A}$] $H_{0}, A$ are  selfadjoint operators, $I$ an open interval and $[0,1)\ni\kappa\mapsto V_{\kappa}$ a symmetric operator valued function such that:
\begin{enumerate}
\item $e^{-isA}D(H_{0})\subset D(H_{0}), \, \forall s\in\bR$, the commutator
$\lbrack A,H_{0}\rbrack$ is $H_{0}$ bounded and a Mourre estimate:
\[\chi(H_{0}\in I) i\lbrack H_{0},A\rbrack\chi(H_{0}\in I)\ge c \chi(H_{0}\in I)+K\]
holds with $c>0$ and $K$ a compact operator; 
\item $ad^{k}_{A}(H_{0})$ are $H_{0}$ bounded for $k\in\{2,3\}$;
\item $ad^{k}_{A}(V_{\kappa})$ are $H_{0}$ bounded  uniformly in $\kappa$ for $k\in\{1,2\}$;
\item  $\kappa\mapsto V_{\kappa} (H_0+i)^{-1}$ is  differentiable in operator norm.
\end{enumerate}
\end{hypo}

\tb{With $P^\perp=\bI-P$, $H^\perp=P^\perp HP^\perp$ and $\left(H^{\perp}-z\right)^{-1}$ its resolvent reduced to $Ran P^\perp$ denote \begin{equation}
F(\kappa,z):=PV_{\kappa}P^{\perp}\left(H^{\perp}-z\right)^{-1}P^{\perp}V_{\kappa}P\quad{Imz\neq0, \kappa\in[0,1)}
\label{def:F}\end{equation}
 the weighted reduced resolvent. The explicit assumptions {\bf(A)} imply the existence of the limit to the real axis  $F(\kappa,E+i0)$ for $E\in I$ and its continuity in operator norm topology, see Theorem  \ref{thm:hs} below; this is in fact all we need to prove the following lower bound on the sojourn time: }

\begin{theorem}\label{thm:main} Assume $({\bf A})$. Let  $\psi$ be an eigenstate of $H_0$ for a simple eigenvalue $E_{0}\in I$. Then it holds for $\kappa\in(0,1)$ and $H=H_{0}+\kappa V_{\kappa}:$
\[\cT(H,\psi)\ge\frac{1}{\Delta E}\]
where the energy width $\Delta E$, which is defined in (\ref{def: energywidth}) below, has the property\\
\[\Delta E=\kappa^{2}\, Im\left\langle\psi,V_0P^{\perp}\left(H_{0}^{\perp}-E_{0}+i0\right)^{-1}P^{\perp}V_0\psi\right\rangle+o(\kappa^{2})\]
and $\cT(H,\psi)=\infty$ if $\Delta E=0$.

Here $P=\ket{\psi}\bra{\psi}$, $P^{\perp}=\bI-P, H_{0}^{\perp}:=P^{\perp}H_{0}P^{\perp}$ and the limit\\
$F(0,E_0+i0)=PV_0P^{\perp}\left(H_{0}^{\perp}-E_{0}+i0\right)^{-1}P^{\perp}V_0P$ exists as a bounded operator.
\end{theorem}

Without attempting to review the huge amount of literature on quantum resonances and time-energy uncertainty principles we make some remarks.

A well known time-energy uncertainty principle is
\[\mathcal{T}\ge\frac{6\pi}{5\sqrt{5}}\frac{1}{\Delta H}\]
with  the uncertainty $\Delta H:=\left(\langle\psi,H^{2}\psi\rangle-\langle\psi,H\psi\rangle^{2}\right)^{1/2}$, see \cite{gsw}. For the situation of Theorem \ref{thm:main} one has $\Delta H=\kappa \Delta{V_\kappa}$ so the denominator of the lower bound is linear in $\kappa$ whereas in Theorem \ref{thm:main} it is quadratic. Also, instead of the uncertainty, our bound involves the term\[Im\left\langle\psi,V_0P^{\perp}\left(H_{0}^{\perp}-E_{0}+i0\right)^{-1}P^{\perp}V_0\psi\right\rangle\] 
which is the overlap of $\psi$ with the continuum eigenstates induced by the perturbation; the occurrence of this term is Fermi's Golden rule. See \cite{rs}, ch XII.6, for more information.

For a general treatment of time-energy principles see Fr\"ohlich--Pfeiffer \cite{pf}.

Lavine's influence on foundations of the theory was important, see \cite{lav1}. Our contribution here is to revisit his ideas with state of the art methods and to relate his lower bound to Fermi's Golden Rule thus making it efficient to produce concrete lower bounds which are in accordance to physics folklore and known mathematical results on related aspects. As pointed out below, see remark \ref{coro:cont}, our theory applies to systems with low regularity meaning that it is sufficient that the extension of $F$ to the real axis is continuous. Furthermore it  is flexible enough to accommodate for systems depending periodically on time. We illustrate this  with two applications to systems which are important for physics and come as a perturbation of an embedded eigenvalue:  the AC Stark effect and  multistate systems, see Sections \ref{sec:stark}, \ref{sec:ms} for additional motivation. With low effort we prove in both cases lower bounds on the sojourn time, equations (\ref{eq:eacs}) and (\ref{eq:ems}), which seem to be new.

The reader can trace back the development of general mathematical resonance theory from the papers of Aguilar Combes \cite{ac}, Simon \cite{s}, Orth \cite{orth}, Cattaneo Graf Hunziker \cite{cgh}. The main focus of these has been on the location of complex poles of an analytic continuation of the resolvent of $H$ across the absolutely continuous spectrum, respectively on exponential decay laws of the form

 \begin{equation}
 \langle\widetilde{\psi},e^{-i t H}\widetilde{\psi}\rangle =a(\kappa)e^{-i t \lambda(\kappa)}+b(\kappa,t)
 \label{eq:edl}\end{equation}
  where the $\widetilde{\psi}$ is related to the  eigenstate $\psi$, $a(\kappa)=1+\cO(\kappa^2)$, \\ 
  $Im(\lambda(\kappa))=\kappa^{2}\, Im\left\langle\psi,F(0,E_0+i0)\psi\right\rangle+o(\kappa^2)$ and $b(\kappa,t)$ can be estimated for  small $\kappa$. 
  
  Our bound on the sojourntime, the square of the $L^2(\bR,dt)$ norm of $\langle\widetilde{\psi},e^{-i t H}\widetilde{\psi}\rangle$, gives complementary information to several points even if the time decay of $b$ is well controlled.  It was proven in  \cite{cgh}, Theorem 1.2 that $\vert b(\kappa, t)\vert\le \frac{c \kappa^2 \vert\log\vert\kappa\vert\vert}{(1+\vert t\vert)}$ under the assumption  that six relatively bounded commutators exist. This estimate works  for $\widetilde{\psi}=g(H)\psi$ with $g$ a smooth function localizing near $E_0$, and  the assumption $Im(F(0,E_0+i0)<0$ was used in a crucial way. The bound on the error $b$ belonging to $L^2(\bR, dt)$, the exponential decay law implies an asymptotic of the sojourn time of $\widetilde\psi$; so in this situation it provides more information than our bound. Here we improve in three points: firstly  we assume only existence of three relatively bounded commutators in ${\bf(A)}.2$ (respectively  two in ${\bf(A)}.3$); secondly  our result concerns the sojourn time of the unperturbed eigenstate $\psi$, a localization of  this state  as in \cite{cgh} is not needed; this is potentially important for applications, see \cite{km} for a discussion of this point related to open systems. Finally our theory covers the case $Im(F(0,E_0+i0))=0$;  remark that an exponential decay law in this case  was proven in estimate (2.19) in \cite{cjn},  which however,  does not provide information on the sojourn time because their error term $b(\kappa,\cdot)$ may not be small in $L^2(\bR,dt)$. The same remark applies to the error bound of \cite{ms}, Theorem 2.1.

Let us mention that, technically speaking, our lower bound on the sojourn time works for systems for which $F(\kappa, E+i0)$ is continuous whereas the results mentioned above need higher order differentiability, see remark \ref{coro:cont} for more information.

Remark that the present method gives lower bounds on the sojourn time. Lower bounds on the Resonance Width, loosely related to Upper bounds on the sojourn time,  where discussed, for example, in \cite{fl,ab}.

In the following section we shall discuss the lower bound in an abstract setting; in Section 3 Theorem \ref{thm:main} is proven in the perturbative situation; in Section 4 we apply to the time periodic case,  see Theorems \ref{thm:mainT} and \ref{thm:stark},  and in Section 5  to multistate systems, see Theorem \ref{thm:mainMS}.

\section{Abstract lower bound on the sojourn time and Lavine's Energy width}

We  review   Lavine's \cite{lav1} abstract lower bound involving the concept of a suitably defined energy width, Theorem (\ref{thm: aprioribound}) below.  Then  we relate it to Fermi's Golden Rule.

\noindent  We denote the resolvent of $H$ by $R(z) = (H - z)^{-1}$.

\begin{lemma}\label{lem:cauchyschwartz}
Let $H$ be a self-adjoint operator on $\cH$. Then it holds for any state $\psi$, any $\lambda \in \bR $ and any
$\epsilon > 0$
\begin{equation}\label{eq:lowerbound}
{\mathcal T}(H,\psi)   \ge  \frac{1}{\epsilon} \, \left(2\epsilon \text{Im} \langle  \psi , \left(H-(\lambda+i\epsilon)\right)^{-1} \psi \rangle \right)^2.
\end{equation}
Equality holds for a $\lambda\in\bR$ and an $\epsilon>0$ if and only if $ \langle  \psi , e^{-iHt} \psi  \rangle = e^{-it \lambda} e^{-\epsilon \vert t\vert}$.
\end{lemma}

{\bf Proof.}
By Fourier Transform, the spectral Theorem and unitarity of $e^{-itH}$ one has
$$
\left\langle\psi,\frac{1}{2i}\, \left(R(\lambda+i\epsilon) - R(\lambda-i\epsilon) \right)\psi\right\rangle = \frac{1}{2}
\int_{\bR} e^{-\epsilon|t| }\, \left\langle\psi,e^{-it(H-\lambda) }\psi\right\rangle dt \, .
$$
Now the claim follows from the Cauchy Schwartz in-equality,
\begin{equation}\label{eq:cs}
\left\vert\text{Im} \langle  \psi , R(\lambda+i\epsilon) \psi \rangle \right\vert\leq  \frac{1}{2}\,
\left( \int_{\bR} e^{-2\epsilon|t| } \right)^{1/2}  ({\mathcal T}(H,\psi))^{1/2}=
\frac{1}{2}\, \frac{\sqrt{{\mathcal T}(H,\psi)}}{\sqrt{\epsilon}},
\end{equation}

where equality holds if and only if $\left\langle\psi, e^{-it (H-\lambda)}\psi\right\rangle=e^{-\epsilon\vert t\vert}$.
\ep

\begin{remarks}\label{rem:equality}
\begin{enumerate}
\item If equality holds in (\ref{eq:lowerbound})  for finite $\epsilon$, then it is a corollary that ${\mathcal T}(H,\psi)  =\frac{1}{\epsilon} $ and thus $ \vert2\epsilon \text{Im} \langle  \psi , R(\lambda+i\epsilon) \psi \rangle\vert=1$.
\item On the other hand, as was remarked in \cite{s}, equality in equation (\ref{eq:cs})  implies analyticity of the spectral measure of $\psi$ and in particular that its support is the whole real line. Thus, equality cannot hold in general, in particular not for semibounded or gapped Hamiltonians.
\end{enumerate}

\end{remarks}
 
Starting from inequality (\ref{eq:lowerbound}) for $\cT$, Lavine defines his notion of energy width.
Given $\lambda\in\bR$, consider the function
\[f(\epsilon) :=   2 \epsilon \,  \text{Im} \langle  \psi , R(\lambda+i\epsilon) \psi \rangle\qquad ( \epsilon>0)\] 

$f$ is non negative,  continuous and monotonous, and $\lim_{\epsilon\to\infty}f(\epsilon)=2$,\quad $\lim_{\epsilon\to0}f(\epsilon)=\tb{2\left\langle \psi, \chi\left(H\in\{\lambda\}\right)\psi\right\rangle}$.

Inspired by   Remarks (\ref{rem:equality}) one defines

\begin{definition}
The energy width of the state $\psi$ at $\lambda\in\bR$ with respect to $H$ is defined as the unique real number
\begin{equation}\label{def: energywidth}
\Delta E:=\Delta E(H, \psi,\lambda):  =
\inf \{\epsilon \in (0,\infty) \, : \,  2 \epsilon \,  \text{Im} \langle  \psi , R(\lambda+i\epsilon) \psi \rangle \ge 1  \}.
\end{equation}

\end{definition}
\begin{remarks}\label{rem:2.4}
\begin{enumerate}
\item \label{rem:bound state}$\Delta E(H, \psi,\lambda)\in\lbrack0,\infty\rbrack$, if $\Delta E(H, \psi,\lambda)=0$ then the projection of $\psi$ on the eigenspace of $H$ at $\lambda$ does not vanish.
\item If $\Delta E(H, \psi,\lambda)>0$ then \[2\Delta E\, Im\left\langle\psi,R(\lambda+i \Delta E)\psi\right\rangle=1\]
\item  An intuition behind the definition of $\Delta E$ is provided by spectroscopy: the energy distribution  of a resonant state at energy $E_{r}$ is expected to be Lorentzian (Cauchy). Testing   this  model, i.e.  calculating $f(\epsilon)$ for $d\langle \psi, E_{\mu} \psi \rangle=\frac{1}{\pi}\frac{\Gamma}{(\mu-E_{r})^{2}+\Gamma^{2}}\, d\mu$,   we get
\[\int_{\bR} \frac{2\epsilon^2}{(\mu - \lambda)^2 + \epsilon^2} \,  \frac{1}{\pi}\frac{\Gamma}{(\mu-E_{r})^{2}+\Gamma^{2}} \, d\mu=\frac{2\epsilon(\epsilon+g)}{(\lambda-E_{r})^{2}+(\epsilon+\Gamma)^{2}}\]
which equals $1$ for $\epsilon=\sqrt{\Gamma^{2}+(\lambda-E_{r})^{2}}$ which in turn is minimal for $\lambda=E_{r}$; in this case $\epsilon=\Delta E=\Gamma$, the half-width at half-maximum of the Lorentzian distribution.
\end{enumerate}
\end{remarks}

With this definition and Lemma (\ref{lem:cauchyschwartz}) one gets

\begin{theorem}\label{thm: aprioribound}
Let $H$ be self-adjoint on $\cH$. Then for any state  $\psi$ in the domain of $H$ and any $\lambda \in \bR$ the following
inequalities hold:
\begin{equation}\label{ETUP}
{\mathcal T}({H,\psi})\geq \frac{1}{ \Delta E(H, \psi,\lambda) }\ge\frac{1}{\Vert (H-\lambda)\psi\Vert }.
\end{equation}
\end{theorem}

{\bf Proof.} The first inequality follows from (\ref{eq:lowerbound}) and the definition of $\Delta E$. For the second inequality it suffices to choose  $\epsilon=\Vert(H-\lambda)\psi\Vert$ and to show that $f\left(\Vert(H-\lambda)\psi\Vert\right)\ge1$ in order to conclude that $\Vert(H-\lambda)\psi\Vert\ge\Delta E$.

 Now, by H\"older's inequality, one has for all $\epsilon>0$
\[1=\int_{\bR}d\mu_{\psi}=\int\sqrt{\frac{(E-\lambda)^{2}+\epsilon^{2}}{\epsilon}}\sqrt{\frac{\epsilon}{(E-\lambda)^{2}+\epsilon^{2}}}d\mu_{\lambda}\le\]
\[\left(\frac{1}{\epsilon}\left(\Vert(H-\lambda)\psi\Vert^{2}+\epsilon^{2}\right)Im\langle\psi,(H-\lambda-i\epsilon)^{-1}\psi\rangle\right)^{1/2}.\]
Taking the square  and choosing $\epsilon=\Vert(H-\lambda)\psi\Vert$ we infer\\ $f\left(\Vert(H-\lambda)\psi\Vert\right)\ge1$.
\ep

In order to relate the energy width $\Delta E$ to Fermi's Golden Rule in a perturbative situation we prove

\begin{propo} Let $H$ be self-adjoint on $\cH$. Let  $\psi$ be a state in the domain of $H$,  $\lambda \in \bR$,  $P$ the orthogonal projector on $\psi$, $\lambda\in\bR$,
$R^{\perp}(z)=\left(H^\perp-z\right)^{-1}$ the resolvent reduced to the subspace $Ran P^{\perp}$. 

If $\Delta E(H,\psi,\lambda)$ defined in (\ref{def: energywidth}) is non zero, then it solves the equation
\begin{equation}
\Delta E=\left\vert \left\langle\psi,\left(H-\lambda-H P^{\perp}R^{\perp}(\lambda+i\Delta E)P^{\perp}H\right)\psi\right\rangle\right\vert.
\label{eq: deltaE}
\end{equation}

\end{propo}

{\bf Proof.} 

Employ ``Feshbach type'' perturbation theory, that is consider $H$ as perturbation of $\mD{H}:=PHP+P^{\perp}HP^{\perp}$, see, for example, \cite{how3}. Denote $R_{\mD}(z):=(\mD{H}-z)^{-1}$.  We have for $PHP^{\perp}$ small enough:
\[PR(z)P=\left(P(H-z)P-PHP^{\perp}R_{\mD}(z)P^{\perp}HP\right)^{-1}\, \mathrm{ and\, thus }\, \]
\begin{equation}\label{eq: Feshbach}
\langle\psi, R(z)\psi\rangle^{-1}= 
\left\langle\psi,(H-z)\psi\right\rangle -F(z)\end{equation}
with
\[F(z):=\left\langle\psi,H P^{\perp}\left(P^{\perp}HP^{\perp}-z\right)^{-1}P^{\perp}H \psi\right\rangle.
\]        
Denoting $\langle  H  \rangle:=   \langle \psi , H \psi  \rangle$ it follows that

\[
2 \Delta E\, Im\langle\psi, R(\lambda+i\Delta E)\psi\rangle=1 \Longleftrightarrow \]
\[ 1=2\Delta E\frac{\Delta E+Im F(\lambda+i\Delta E)}{\left(\langle H\rangle-\lambda-Re F(\lambda+i\Delta E)\right)^{2}+\left(\Delta E+Im F(\lambda+i\Delta E)\right)^{2}}\]
\[\Longleftrightarrow \Delta E^{2}= \left(\langle H\rangle-\lambda-Re F\right)^{2}+Im F^{2}=\left\vert\langle H\rangle-\lambda-F\right\vert^2.\]
\ep\\

\section{Quadratic lower bound for a perturbed embedded eigenvalue}
\label{sec:qlb}

We now prove the lower bound on the sojourn time  ${\mathcal T}(H,\psi)$ stated in Theorem \ref{thm:main}. 
 We  make use of a limiting absorption principle and continuity properties of the boundary values of the resolvent on states orthogonal to the unperturbed bound state which are known to hold under hypothesis ${\bf(A)}$, see \cite{hs,cgh,mw}; the general idea is to show that a strong Mourre estimate holds for the reduced operator $H^\perp$ and then to apply the classical results in the spirit of \cite{jmp}. We first recall the meaning of the commutators $ad^k_A(H_0)$ in hypothesis {\bf (A)}; this construction, originally developed in \cite{m,jmp}  starts from 
 the sesquilinear form $i (H_0u, Av) - i (Au, H_0v)$  which is well defined
for all vectors $u$, $v$ in the domain $\mathcal{D}(H_0)\cap\mathcal{D}(A)$. 
We quote the efficient resum\'e of  \cite{cgh} made in the paragraph following their equation $(4)$:

\setlength{\leftskip}{.3cm}

{If this form has a bound 
\begin{equation}
 |i(H_0u, Av) - i(Au, H_0v)|\leqslant C\|u\|\|(H_0+i)v\|\,,
\label{bd}
\end{equation}
it extends to the sesquilinear form of a unique self-adjoint operator 
called $i[H_0, A]=i ad_A^1(H_0)$ which is bounded relative
to $H_0$. Therefore the second order commutator 
$i[i[H_0, A],A]= -\mathrm{ad}^{2}_A(H_0)$ is defined as well if the bound 
(\ref{bd}) with $i[H_0, A]$ instead of $H_0$ on the left hand side is assumed. 
For $k\ge3$ the $k$-th order commutator, denoted by $\mathrm{ad}^{k}_A(H_0)$, 
is then defined recursively in terms of $\mathrm{ad}^{k-1}_A(H_0)$ and $A$.}

\setlength{\leftskip}{0pt}

\noindent Recall the definition of the weighted reduced resolvent (\ref{def:F}). We have

\begin{theorem}\label{thm:hs}
Assume ${\bf (A)}$ and let $E_{0}$ be a simple eigenvalue of $H_{0}$ with eigenprojector $P$. Then there exists an interval $I\ni E_{0}$ and a $\kappa_{0}>0$ such that $\forall\kappa, \vert\kappa\vert<\kappa_{0}$, $\forall\lambda\in I$ the norm limit 
\[F(\kappa,\lambda+i0)=\lim_{\epsilon\searrow0}PV_{\kappa}P^{\perp}\left((H_{0}+\kappa V_{\kappa})^{\perp}-(\lambda+i\epsilon)\right)^{-1}P^{\perp}V_{\kappa}P\]
exists and is bounded uniformly in $\vert\kappa\vert<\kappa_{0}$ and $\lambda\in I$;
furthermore for $\vert\kappa\vert<\kappa_{0}$, $Re(z)\in I$
\[\Vert F(\kappa,z)-F(\kappa^{\prime},z^{\prime})\Vert\le c \left(\vert\kappa-\kappa^{\prime}\vert+\vert z-z^{\prime}\vert\right)^{\frac{1}{3}}\]
\end{theorem}

{\bf Proof}. The result is proven in \cite{hs}, Lemma $8.11$ under the assumption that $\psi\in D(A^2)$  and that two relatively bounded  commutators exist. In \cite{mw} it is proven that the relative boundedness of $ad_A^3(H_0)$  implies  $\psi\in D(A^2)$.
\ep

\begin{remark}
Notice that in Theorem \ref{thm:main} Assumption {\bf (A)}.2 may be replaced by:
\begin{center}
 $ad^{2}_{A}(H_{0})$ is $H_{0}$ bounded and the eigenvector $\psi$ belongs to $D(A^2)$.
\end{center}
Note that \cite{mw} constructed an example with two commutators relatively bounded and $\psi\notin D(A^2)$.
\end{remark}

We now finish the proof of our main result.

{\bf Proof of Theorem (\ref{thm:main}).}  Suppose $\Delta E>0$, otherwise  $\mathcal{T}$ is infinite, see  Remark \ref{rem:2.4}.\ref{rem:bound state}.  Denote the resolvent with respect to $H_{0}+\kappa V_{\kappa}$ by $R_{\kappa}$ and $\langle  A \rangle_{\varphi}:=\langle\varphi, A\varphi \rangle$. We take account of the Lamb shift, meaning that  in equation (\ref{eq: deltaE}) we choose 
 \[\lambda=\lambda_{2}(\kappa):=E_{0}+\kappa\langle V_{\kappa}\rangle_{\psi}-\kappa^{2}Re\left\langle R_{0}^{\perp}\left(E_{0}+i0\right)\right\rangle_{P^{\perp}V_{\kappa} P\psi}.\]
 
 Thus
 \begin{eqnarray*}
&&\Delta E(\lambda_{2}(\kappa))=\\
&&\kappa^{2}\left\vert Re\left\langle R_{\kappa}^{\perp}\left(\lambda_{2}+i\Delta E\right)- R_{0}^{\perp}\left(E_{0}+i0\right)\right\rangle_{P^{\perp}V_{\kappa} P\psi}+i Im\left\langle R^{\perp}\left(\lambda_{2}+i\Delta E\right)\right\rangle_{P^{\perp}V_{\kappa} P\psi}\right\vert.
\end{eqnarray*}

From  Theorem \ref{thm:hs} we conclude firstly that $\Delta E(\lambda_{2}(\kappa))=\cO(\kappa^{2})$ and secondly that

\[\left\vert \Delta E(\lambda_{2}(\kappa))- 
\kappa^{2}\, Im\left\langle R_{0}^{\perp}\left(E_{0}+i0\right)\right\rangle_{P^{\perp}V_{\kappa} P\psi}\right\vert
\le
c\kappa^{2}\left(\kappa+\Delta E+\vert\lambda_{2}-E_{0}\vert\right)^{\frac{1}{3}}=\cO\left(\kappa^{\frac{7}{3}}\right).\] 
\ep
\tb{\begin{remark}It is crucial that the a priori estimate (\ref{ETUP}) is valid for any parameter $\lambda$. So we can choose $\lambda_2(\kappa)$ above and thus implement the intuition provided by other theories. For example it was shown in \cite{cgh}, Theorem 1.2,  that in the exponential decay law (\ref{eq:edl}) it holds 
$Re(\lambda(\kappa))=\lambda_2(\kappa)+o(\kappa^2)$.  The Lamb shift is determined by the perturbations  $Re(E(\kappa))-E_0$ where $E_0$ stand for a degenerate eigenvalue of the hydrogen atom and $E(\kappa)$ for the  resonance induced from $E_0$ by the interaction of the atom with a quantized electromagnetic field. 
By calling the real quadratic correction to $E_0$  "Lamb shift"   in the proof of the theorem above we refer to an important achievement of mathematical physics:  the proof that, for a suitable model, the second order contribution to the Lamb shift is determined by $\kappa^2 Re F(0,E_0+i0)$, see \cite{bfs}, Theorem I.3.
\end{remark}
}
\tb{
\begin{remark}\label{coro:cont}
The proof actually implies the bound $\cO(\kappa^{7/3})$ for the error; only $o(\kappa^2)$ is stated in Theorem \ref{thm:main}. Remark that  the continuity of the function $(\kappa,\lambda)\mapsto F(\kappa, \lambda+i0)$  is sufficient to  proof of Theorem \ref{thm:main}. So instead of {\bf (A)} we could have assumed this property; we opted for hypothesis {\bf (A)} because it is simple and explicit. Remark, however,  that the (H\"older-) continuity of $F$, can be inferred by other methods, for example, from singular Mourre theory, see \cite{fms}, Theorem 3.3. The above mentioned works on exponential decay rules, \cite{ms, cgh, cjn} assume higher order differentiability.
\end{remark}
}
\section{Time periodic perturbations}\label{sec:stark}
\tb{One feature of the simplicity of our theory is that  applies immediately to  time periodic perturbations of Schr\"odinger operators via Floquet theory. A particular  special case is the two body AC Stark effect which is maybe the most simple physically relevant system to which our theory applies. Location and exponential decay  for resonances were analyzed in detail by Yajima and Graffi  \cite{y,gy} using complex deformation methods and by M\o ller and Skibstedt \cite{ms} in great generality using Mourre techniques. Here we  aim only at lower bounds on the sojourn time in the two body case for smooth potentials which were not discussed before.
}

Consider for $t\in\bR$
\[H_{0}=-\frac{1}{2}\Delta +W\qquad \hbox{ and }\qquad H(t)=H_{0}+\kappa V_{\kappa}(t)\qquad \hbox{ on } H^{2}(\bR^{d});\]
We assume that that   $T-$periodic family $V_{\kappa}(t)$ and $W$ as a constant function of $\kappa, t$ satisfy:

\begin{hypo}[${\bf AT}$] Let $(\kappa, t, x)\mapsto g_{\kappa}(t,x) \in C^{\infty}(\lbrack0,1)\times\bR/(T\bZ)\times\bR^{d}; \bR)$ be such that for a $\delta>0$ and all $\alpha\in\bN_{0}^{d}$
\[\sup_{\kappa\in[0,1),t\in[0,T]}\sup_{x\in\bR^d}\Vert\langle x\rangle^{\delta+\alpha}\partial_x^{\alpha}g_{\kappa}(t,x)\Vert<\infty.\]\end{hypo}

With $\omega:=\frac{2\pi}{T}$ we denote by 

\[\widehat{V_{\kappa}}(n)=\frac{1}{{T}}\int_{0}^{T}\exp{(-i n\omega t)}V_{\kappa}(t)\ dt\]
the multiplication operator in $L^2(\bR^d)$ by the $n-$th Fourier coefficient of $V_{\kappa}$. Remark that $\widehat{V_{\kappa}}(n)$=$\overline{\widehat{V_{\kappa}}(-n)}$.

Under assumption $({\bf AT})$ the propagator  $U$ generated by $H(\cdot)$ is well defined; the sojourn time depends on the initial time $t_{0}$

\[{\mathcal T}(H(\cdot),\psi, t_{0})=\int_{-\infty}^{\infty}\left\vert\left\langle\psi, U(t,t_{0})\psi\right\rangle\right\vert^{2} \ dt.\]
We now prove a lower bound for its average over $t_0$:

\begin{theorem}\label{thm:mainT} Suppose that $W$ and $V_\kappa$ satisfy  assumption ${\bf (AT)}$.  Let $H_{0}=-\frac{1}{2}\Delta+W$ on $H^2(\bR^d)$. Let $\psi$ be a simple eigenstate of $H_0$ with  eigenvalue $E_{0}\in I\subset\bR\setminus\omega\bZ$ which is non-resonant, i.e.:  $E_0+\omega\bZ\bigcap\sigma_{pp}(H_0)=\{E_0\}$.  Then it holds for the lifetime of $\psi$ with respect to $H(\cdot)$ averaged over the initial time $t_0$:

\[\frac{1}{T}\int_{0}^{T}\cT(H(\cdot),\psi,t_{0})\ dt_{0}\ge\frac{1}{\Delta E}\]
where the energy width $\Delta E$ has the property

\[\Delta E=\kappa^{2}\,\sum_{n\in\bZ} Im\left\langle\psi,{\widehat{V_0}(-n)}\widetilde{R}_0\left(E_0+n+i0\right)\widehat{V_0}(n)\psi \right\rangle+o(\kappa^{7/3}).\]
Here $\widetilde{R}_0\left(E_0+n+i0\right)$ is the limit of the resolvent for $H_0$ for $n\neq0$ and the reduced resolvent for $n=0$,  $\widehat{V_0}(n):=\frac{1}{T}\int_0^T e^{in\omega t}V_0(t)\ dt$.\end{theorem}

To prove Theorem (\ref{thm:mainT}) we use the space-time picture advocated by Howland, see \cite{how1}, and apply our theory to the Floquet Hamiltonian
\[K:=K_{0}+\kappa V_{\kappa}(t), \quad K_{0}:=-i\partial_{t}-\Delta+W\qquad\hbox{ on } H^{1}(\bT)\otimes H^{2}(\bR^{d})\]
$\bT:=\bR/(T\bZ)$. It is known, \cite{how1} formula 1.6,  that for ${\boldsymbol\varphi}\in L^{2}(\bT)\otimes L^2(\bR^{d})$:
\begin{equation}\label{eq:howland}
\left(e^{-iKs}{\boldsymbol\varphi}\right)(t+s)=U(t+s,t){\boldsymbol\varphi}(t).
\end{equation}
Denote for $n\in\bZ$ the harmonics $ h_{n}(t):=\frac{1}{\sqrt{T}}\exp{(i n\omega t)}$ and $P_{n}:L^{2}(\bT)\to L^{2}(\bT)$ the projector on $span\{ h_{n}\}$.  $-i\partial_t$ and $H_0$ commute so it holds for $z\in\bC\setminus\bZ$ strongly:
\begin{equation}
\left(K_{0}-z\right)^{-1}=\sum_{n\in\bZ}P_{n}\otimes\left(H_{0}+n-z\right)^{-1}.\label{floquetresolvent}
\end{equation}

{\bf Proof} (of Theorem (\ref{thm:mainT})).
In \cite{yoko}, Theorem 1.3 and Lemma 2.4, it was proved that  {\bf(AT)} implies selfadjointness of $K_{0}, K$ 
and  that {\bf (A)} holds in a neighborhood  $I$ of $E_0$ away from $\omega\bZ$ for the quadruple \quad $K_{0}, \bV_{\kappa} , \bA, I$   with $\bV_{\kappa} $ the operator valued multiplication by $V_{\kappa} (t)$,  $\bA:=\bI\otimes \frac{1}{2}(xL_D+L_D x)$ and $L_D:=D(D^2+1)^{-1}, D:=-i\nabla$.  

 The non-resonance condition ensures that $E_{0}$ is a simple eigenvalue of $K_{0}$ with eigenstate $h_{0}\otimes\psi$. Application of Theorem (\ref{thm:main}) to the operator $K$ yields

\[\mathcal{T}(K,h_{0}\otimes\psi)\ge\frac{1}{\Delta E}\]
with
\[\Delta E=\kappa^2 Im\left\langle h_{0}\otimes\psi,\bV_0\bP^{\perp}\left(K_{0}^{\perp}-E_{0}+i0\right)^{-1}\bP^{\perp}\bV_0 h_{0}\otimes\psi\right\rangle_{L^{2}(\bT)\otimes L^{2}(\bR^{d})}+o(\kappa^{2})\]

for $\bP=\ket{h_{0}\otimes\psi}\bra{h_{0}\otimes\psi}$. Now $\bP^{\perp}=P_{0}^{\perp}\otimes\bI+P_{0}\otimes P_{\psi}^{\perp}$ so by (\ref{floquetresolvent}):
\[\left(K_{0}^{\perp}-z\right)^{-1}=\sum_{n\neq0}P_{n}\otimes\left(H_{0}+n-z\right)^{-1}+P_{0}\otimes\left(H_{0}^{\perp}-z\right)^{-1}.\]
Furthermore $P_0\otimes P_\psi\bV_0=\sum_m\ket{h_0}\bra{h_m}\otimes P_\psi\widehat{V_0}({0-m})$ which implies
\[\Delta E=\kappa^{2}\,\sum_{n\in\bZ} Im\left\langle\psi,{\widehat{V_0}(-n)}\widetilde{R}_0\left(E_0+n+i0\right)\widehat{V_0}(n)\psi \right\rangle+o(\kappa^{7/3}).\]

For the case $\frac{1}{T}\int_{0}^{T}\cT(H,\psi,t_{0})\ dt_{0} <\infty$ the result now follows  from Jensen's inequality and Fubini's Theorem:
\begin{eqnarray*}
\cT(K,h_{0}\otimes\psi)&=&\int_{\bR}\left\vert\frac{1}{T}\int_{0}^{T}\left\langle\psi,U(t_{0}+s,t_{0})\psi\right\rangle\ dt_{0}\right\vert^{2} ds\\
&\le&\int_{\bR}\frac{1}{T}\int_{0}^{T}\left\vert\left\langle\psi,U(t_{0}+s,t_{0})\psi\right\rangle\right\vert^{2}\ dt_{0}\ ds\\
&=&\frac{1}{T}\int_{0}^{T}\cT(H(\cdot),\psi,t_{0})\ dt_{0}
\end{eqnarray*}\ep

We now apply this general estimate to the AC Stark effect and obtain
\begin{theorem}\label{thm:stark}
Let  $W\in C^{\infty}(\bR^{d}, \bR)$  such that for a $\delta>0$ and $\alpha\in\bN_{0}^{d}$
$\sup_{x}\Vert\langle x\rangle^{\delta+\alpha}\partial_x^{\alpha}W(x)\Vert<\infty$.  Let $F\in C^{\infty}(\lbrack0,T\rbrack;\bR^d)$, be a  $T$-periodic function with zero mean and Fourier series $F(t)=\sum_{n\neq0}F_n e^{i n \omega t}$. 

Let $q\in C^\infty(\bR; \bR^d)$, $q(t)=\sum_{n\neq0}\frac{F_n}{(in\omega)^2} e^{i n \omega t}$ which is a $T$ periodic function such that $\ddot q=F$.

Let $E_{0}\in\bR\setminus\omega\bZ$ be a simple  eigenvalue of $H_{0}:=-\frac{1}{2}\Delta+W(x)$ on $H^{2}(\bR^{d})$ with eigenvector $\psi$ such that $E_0+\omega\bZ\bigcap\sigma_{pp}(H_0)=\{E_0\}$.

For the sojourn time of $\psi$ with respect to the propagator generated by 
\[H(t):=-\frac{1}{2}\Delta+W(x+\kappa q(t)) \qquad \hbox{ on } H^{2}(\bR^{d}), \quad \kappa\in\lbrack0,1),t\in\bR :
\]
it holds:
\[\frac{1}{T}\int_{0}^{T}\cT(H,\psi,t_{0})\ dt_{0}\ge\frac{1}{\Delta E}\]
where the energy width $\Delta E$ satisfies, 

\begin{equation}
\Delta E=\kappa^{2}\,\sum_{n\in\bZ} Im\sum_{j,k=1}^d\frac{\overline{F_n^j}F_n^k}{\omega^4 n^4}\left\langle\psi,{\partial_j W}\widetilde{R}_0(E_0+n+i0)\partial_k W\psi \right\rangle+o(\kappa^{7/3})\label{eq:eacs}
\end{equation}
and $\widetilde{R}_0\left(E_0+n+i0\right)$ denotes the limit of the resolvent of $H_0$ if $n\neq0$ and the reduced resolvent if $n=0$.\end{theorem}

\begin{remark}$H(\cdot)$ is the Hamiltonian of an electron in the potential $W$ and the homogeneous electric field of strength $\kappa F(\cdot)$ in the frame of a freely falling observer. To see this, consider the unitary family 
of phase space translation operators $S(t)$ on   $L^2(\bR^d)$
\[S(t) = e^{-i\left(q(t)D-p(t)x\right)}e^{i\varphi(t)}\]
with $T$ periodic functions $q,p$ such that $\dot{p}=\kappa F$ and $\dot{q}=p$ and $\varphi\in C^\infty(\bR,\bR)$ such that $\dot{\varphi}=\frac{\dot{q}p-\dot{p}q}{2}-\left(\frac{p^2}{2}-\kappa F q\right)$.
Now from the identities 
\[S^{-1}xS=x+q, S^{-1}DS=D+p\]
\[ i\partial_t e^{-i\left(q(t)D-p(t)x\right)}\psi=\left(\left(\dot{q}D-\dot{p}x\right)+\frac{1}{2}\left(\dot{p}q-\dot{q}p\right)\right)e^{-i\left(q(t)D-p(t)x\right)}\psi\]
and  $\psi\in\cS(\bR^d)$ it follows that
\[S^{-1}\left(-i\partial_t+\frac{D^2}{2}-\kappa F x+W(x)\right)S\psi=\left(-i\partial_t+\frac{D^2}{2}+W(x+\kappa q(t))\right)\psi.\]
Thus if $\psi(t)$ solves the Schr\"odinger equation for $\frac{D^2}{2}+W\left(x+\kappa q(t)\right)$ then $S(t)\psi(t)$ solves the Schr\"odinger equation for $\frac{D^2}{2}-\kappa F x+W(x)$ in the laboratory frame.
\end{remark}

{\bf Proof} (of Theorem (\ref{thm:stark})).

Define $V_{\kappa}(t)$ by
\[\kappa V_{\kappa}(t,x):=W(x+\kappa q(t))-W(x).\]
Then $V_{\kappa} $ satisfies Hypotheses {\bf(AT)} and the sojourn time estimate stated in Theorem \ref{thm:mainT} holds. Now $V_0(t,x)=\left(\partial_\kappa W\right)(x)=\sum_{j=1}^d q_j(t)\partial_jW(x)$ and 
\[\widehat{V_0}(x)=\sum_j\frac{F_n^j}{(in\omega)^2}\partial_jW(x)\]
from which the formula for $\Delta E$ follows.\ep

\section{Sojourn time of multistate systems}\label{sec:ms}

Systems with inner degrees of freedom appear in various physical situations. We just mention the Dirac equation and effective Hamiltonians which appear in the Born Oppenheimer approximation, see \cite{cs,hj,t}. Here we are interested in situations where one channel is binding and the others are propagating; this occurs, for example, in molecular predissociation, \cite{k, mg}.

\begin{theorem}\label{thm:mainMS} Consider self-adjoint operators two Hilbert spaces $\cH_1$ and $\cH_2$ and  in $\cH_1\oplus\cH_2$
\[
H:=\left(
\begin{array}{cc}
  H_1   & \kappa V_{\kappa}  \\
 \kappa V_{\kappa}^\ast &   H_2  
\end{array}
\right)
\]
such that:

$H_1$ is self-adjoint in $\cH_1$,  $E_0$ is a simple eigenvalue of $H_1$ with normalized eigenvector $\psi_0$ and  the resolvent if $H_1$ is compact in a punctured complex neighborhood of $E_0$;

$H_2$ is self-adjoint in $\cH_2$,  there exists a selfadjoint $A$ in $\cH_2$ and an interval $I$ around $E_0$ such that $e^{-isA}D(H_2)\subset D(H_2)$ for $s\in\bR$ and $ad_A^k(H_2)$ is $H_2-$bounded for $k\in\{1,2\}$ and such that for a positive number $c$ it holds
\[\chi(H_2\in I)i\lbrack H_2,A\rbrack\chi(H_2\in I)\ge c \chi(H_2\in I);\]

the values of  $\lbrack0,1)\ni\kappa\to V_{\kappa}$ are operators from $\cH_2$ to $\cH_1$ such that
\[
\kappa\mapsto V_{\kappa}\left(H_2+i\right)^{-1} \quad\hbox{ \rm and } \kappa\mapsto V^\ast_\kappa\left(H_1+i\right)^{-1}
\]
are norm differentiable  and such that 
\[A^kV_{\kappa}^\ast\left(H_1+i\right)^{-1} \quad\hbox{ \rm and }V_{\kappa} A^k\left(H_2+i\right)^{-1}\]
extend to bounded operators for $k\in\{1,2\}$. Then it holds for 
$H$ :

\[\cT(H,\psi_0\oplus 0)\ge\frac{1}{\Delta E}\]
where  the energy width $\Delta E$ has the property

\begin{equation}
\Delta E=\kappa^{2}\, Im\left\langle\psi_0,V_0\left(H_2-(E_0+n)+i0\right)^{-1}V_0^\ast\psi_0 \right\rangle+o(\kappa^{7/3}).\label{eq:ems}
\end{equation}
\end{theorem}

{\bf Proof} (of Theorem (\ref{thm:mainMS})).
Define 
\[H_0:=H_1\oplus H_2,\quad \bA:=0\oplus A,\quad \bV_{\kappa}:=\left(\begin{array}{cc}0&V_{\kappa}\\V_{\kappa}^\ast&0\end{array}\right).\]
 We  show that the quadruple $H_0, \bV, \bA, I$ satisfies assumption {\bf(A)} and apply Theorem \ref{thm:main}. Indeed:

$e^{-i s \bA}=\bI\oplus e^{-isA}$ leaves $D(\bH_0)$ invariant by the assumption on $A$; also $ad_\bA^k(\bH_0)=0\oplus ad_A^k(H_2)$ are $\bH_0$ bounded. 
\begin{eqnarray*}
 \chi(\bH_0\in I)i \lbrack\bA,\bH_0\rbrack\chi(\bH_0\in I)&=&0\oplus\chi(H_2\in I)i\lbrack A,H_2\rbrack\chi(H_2\in I)\\
\ge c 0\oplus\chi(H_2\in I)&=&c\chi(\bH_0\in I)- c\chi(H_1\in I)\oplus 0
\end{eqnarray*}
by the Mourre estimate for $H_2$, the second term is compact, upon shrinking $I$ if necessary, because the resolvent of $H_1$ near $E_0$ is compact. Furthermore for $k\in\{0,1,2\}$
\[ad_\bA^k(\bV_{\kappa})=\left(\begin{array}{cc}0& (-1)^k V_{\kappa}A^k\\ A^kV_{\kappa}^\ast&0\end{array}\right)\]
is relatively $\bH_0$ bounded. Remark that $\bA^k$ acts on the $\cH_2$ component so $\psi_0\oplus 0$ is in $D(\bA^k)$ and we have proven that the assumption {\bf(A)} is satisfied. Now the claim follow from Theorem \ref{thm:main} with the observation 
\[P^\perp=\left(\begin{array}{cc}\bI- \ket{\psi}\bra{\psi}& 0\\ 0&\bI\end{array}\right),\quad P^\perp\bV_{\kappa} P= \left(\begin{array}{cc} 0&0 \\ \ket{V_{\kappa}^\ast\psi}\bra{\psi}&0\end{array}\right).\]\ep

\bigskip
{\bf Acknowledgements}
We   gratefully acknowledge support from the
grants Fondecyt Grants 1120786 and 1141120;  Anillo  Conicyt PIA-ACT1112, Ecos-Conicyt C10E01, EPLANET. OB, VC, CF thank CPT and JA thanks Facultad de Matem\'aticas of PUC for hospitality.


\begin{thebibliography}{cexx}

\bibitem{ac} Aguilar, J. and Combes, J.M. {\it A class of analytic perturbations for one-body Schr\"odinger Hamiltonians}. Comm. Math. Phys. {\bf 22} (1971), 269--279.

\bibitem{ab}Asch, J. and Briet, P.,  {\it Lower bounds on the width of Stark-Wannier type resonances}. Comm.Math.Phys., (1996) 179(3), pp.725--735.


\bibitem{bfs}Bach, V., Fr\"ohlich, J., Sigal, I.M., 1998. Quantum electrodynamics of confined nonrelativistic particles. Advances in Mathematics, 137(2), pp.299--395.

\bibitem{cgh}
 Cattaneo, L., Graf, G. M., and Hunziker, W., {\it A general resonance
theory based on Mourre's inequality}, Ann. Henri
Poincar\'e, 7 (2006),  583--601.

\bibitem{cs}Combes, J.M. and Seiler, R. {\it The Born Oppenheimer approximation}. In: Rigorous atomic and molecular physics. Velo, G. Wightman, A. (eds.) pp. 185--212. New York: Plenum Press 1981.





\bibitem{cjn}Cornean, H.D., Jensen, A.,  Nenciu, G., 2014. Metastable States When the Fermi Golden Rule Constant Vanishes. Communications in Mathematical Physics, 334(3), pp.1189--1218.

\bibitem{fms}Faupin, J., M\o ller, J.S., Skibsted, E., 2011. Second order perturbation theory for embedded eigenvalues. Communications in Mathematical Physics, 306(1), pp.193--228. 

\bibitem{fl} Fernandez, C. and Lavine, R., {\it Lower bounds for resonance widths in potential and obstacle scattering}. Comm. Math. Phys., (1990), 128(2), pp.263--284.

\bibitem{gsw} {Gislason, Eric A and Sabelli, Nora H and Wood, John W},
 {{\it New form of the time-energy uncertainty relation}},
 {Physical Review A},  {31}, {4}, {(1985)}, {2078--2081}
 

\bibitem{mg} Grigis, A. and Martinez, A. {\it Resonance widths for the molecular predissociation}. Anal. PDE 7 (2014), no. 5, 1027--1055
 
\bibitem{gy} Graffi, S. and Yajima, K.,  {\it Exterior complex scaling and the AC-Stark effect in a Coulomb field.} Communications in Mathematical Physics, 89(2), pp. 277 -- 301, 1983.

\bibitem{hj} Hagedorn, J. and Joye, A. {\it Mathematical Analysis of Born-Oppenheimer Approximations}, in 
Spectral Theory and Mathematical Physics: A Festschrift in Honor of Barry Simon's 60th Birthday. F. Gesztesy, P. Deift, C. Galvez, P. Perry, W. Schlag, editors. 
AMS Proc. of Symposia in Pure Math. 76 , p. 203-226, 2007.

\bibitem{how1}
Howland, J. , {\it Stationary scattering theory for time-dependent
Hamiltionian,} Math. Ann., 207 (1974), 315-335.



\bibitem{how3}
Howland, J. , {\it The Livsic matrix in perturbation theory},
 {Journal of Mathematical Analysis and Applications}, Vol. {50}, {1975}, {415--437}
 
 \bibitem{hs} {Hunziker, W and Sigal, I. M.},
{{\it The quantum $N$-body problem}},
{Journal of Mathematical Physics},
{Vol. 41},
(2000),
{3448--3510}

\bibitem{jmp}Jensen, A., Mourre, E. and Perry, P., 1984. Multiple commutator estimates and resolvent smoothness in quantum scattering theory. Annales de l'Institut Henri Poincar\'e. Section A. Physique Th\'eorique. Nouvelle S\'erie, 41(2), pp.207--225.

  \bibitem{k} Klein, M. {\it On the mathematical theory of predissociation}. Ann. Physics 178 (1987), no. 1, 48--73. 
  
  \bibitem{km} K\"onenberg, M., Merkli, M., 2015. On the irreversible dynamics emerging from quantum resonances, \url{http://arxiv.org/pdf/1503.02972.pdf}
  
\bibitem{lav1}
Lavine, R., {\it Spectral density and sojourn times, Atomic Scattering Theory } (J.Nutall, ed.), U.
of Western Ontario, London, Ontario, 1978.




\bibitem{mims}Merkli, M. and Sigal, I.M. , {\it A time-dependent theory of quantum resonances.} Communications in Mathematical Physics, 201(3), pp.549--576,  1999.

\bibitem{ms} M\o ller, J.S. and Skibsted, E.,  Spectral theory of time-periodic many-body systems. Advances in Mathematics, 188(1), pp.137-- 221, 2004.

\bibitem{mw}M\o ller, J.S. and Westrich, M.,  Regularity of eigenstates in regular Mourre theory. Journal of Functional Analysis, 260(3), pp.852--878, 2011.

\bibitem{m}Mourre, E., {\it Absence of singular continuous spectrum for certain selfadjoint operators.} Communications in Mathematical Physics, 78(3), pp.391--408, (1981).

\bibitem{orth}
Orth, A. , {\it Quantum mechanical resonance and limiting absorption: the many body problem},
 {Communications in Mathematical Physics}, Vol. {126}, {1990}, {559--573}
 
\bibitem{pf} {Pfeifer, Peter and Fr{\"o}hlich, J{\"u}rg},
 {{\it Generalized time-energy uncertainty relations and bounds on lifetimes of resonances}}. {Reviews of Modern Physics},  {67}, {4},  ({1995}),  {759--779}



\bibitem{rs} Reed, M. and Simon,  B. , {\it Methods of Modern Mathematical Physics}, Vol.1-4, Academic Press, New York, 1975-1979.

\bibitem{y} Yajima, K., {\it Resonances for the AC-Stark effect.} Communications in Mathematical Physics, 87(3), 331 -- 352, 1982
\bibitem{yoko} Yokoyama,  K., {\it Mourre Theory for Time-periodic Systems}, Nagoya Math. J., {\bf 149} (1998), 193-210.

\bibitem{s} {Simon, Barry},
 {{\it Resonances and complex scaling: a rigorous overview}},
 {International Journal of Quantum Chemistry}, {14},{4},{(1978)}, {529--542}
\bibitem{skib}
Skibsted, E., {\it Truncated Gamow functions, $\alpha$--decay and the exponential Law},
Commun. Math. Phys. 104, 591--604, 1986.

\bibitem{t}Teufel, S. {\it Adiabatic perturbation theory in quantum dynamics}. Lecture Notes in Mathematics, 1821. Springer-Verlag, Berlin, 2003







\end{thebibliography}
  \end{document}